\documentclass[twocolumn,amsmath,amssymb,aip,pop]{revtex4}

\usepackage{graphicx}
\usepackage{bm}

\usepackage{amsfonts}

\begin{document}
\title{Short-wavelength soliton in ultrarelativistic electron-positron-ion plasmas}
\author{V. M. Lashkin$^{1,2}$}
\email{vlashkin62@gmail.com} \affiliation{$^1$Institute for
Nuclear Research, Pr. Nauki 47, Kyiv 03028, Ukraine}
\affiliation{$^2$Space Research Institute, Pr. Glushkova 40 k.4/1,
Kyiv 03187,  Ukraine}

\begin{abstract}
We derive a nonlinear equation governing dynamics of
short-wavelength longitudinal waves in ultrarelativistic
electron-positron-ion plasmas. In contrast to the recent work by
Lashkin [Phys. Plasmas {\textbf{27}}, 102302 (2020)], where a
similar equation was suggested in the framework of the Wigner
function approach for a nonrelativistic electron-ion degenerate
plasma, in our case which is based on the Vlasov kinetic equation
all three species of particles (electrons, positrons and ions)
should be present. The nonlinearity arises only in the presence of
a population of ions. By numerical simulations we demonstrate that
collisions between even four solitons are fully elastic.
\end{abstract}

\maketitle

Ultrarelativistic  plasmas exist in various astrophysical objects
such as supernova remnants, pulsars, active galactic nuclei etc.
\cite{Raffelt96,Thoma09}, and also provide important insights
about the early stage in the evolution of Universe
\cite{Weinberg_book,Misner1980}. In laboratory conditions such
plasmas can be produced in high-intensity laser fields
\cite{Shen01,Liang08}. Such plasmas are always arises if the
thermal energy of the particles exceeds twice rest mass energy of
electrons $\sim 1.2$  MeV \cite{Tajima1990}. Most astrophysical
and laboratory ultrarelativistic plasmas consist of electrons,
positrons and a minority population of ions (typically, the latter
are nonrelativistic).  Nonlinear waves and solitons in
relativistic plasmas have been studied extensively for the past
four decades (see for example a review \cite{Shukla1986}).
Specifically in an nondegenerate ultraretivistic plasma, nonlinear
evolution equations and their soliton solutions have been
considered in a number of works. As was shown in Refs.
\cite{Mofiz1992,Mofiz1993}, dynamics of ultrarelativistic Langmuir
waves in electron-positron plasma in the framework of the fluid
model is governed by the nonlinear Schrodinger (NLS) equation and
the corresponding solution is the ultrarelativistic Langmuir
soliton, and, in particular, its relevance to pulsar radiation was
discussed. Later on, the NLS equation and ultrarelativistic
Langmuir soliton in electron-positron plasma were obtained from
the kinetic approach based on the Vlasov equation with
ultrarelativistic Maxwellian distribution function \cite{Liu2012}.
Ultrarelativistic Alfv\'{e}n solitons propagating parallel and
oblique to the external magnetic field correspond to the
Korteweg-de Vries (KdV) equation and were studied in Refs.
\cite{Sakai1980} and \cite{Verheest1997} respectively. The
considered models are valid only in the long-wavelength case $k\ll
1$, where $k$ is suitably normalized dimensionless wavenumber.
Then, the appearance of solitons is due to the balance of weak
dispersion $k^{2}\ll 1$ for the NLS equation, and  $k^{3}\ll 1$
for the KdV equation, and weak nonlinearity (cubic and quadratic
respectively). Recently, ion-acoustic solitons in a plasma with
ultrarelativistic electrons and positrons  were studied in detail
in Ref. \cite{Verheest2019} using the fluid model and the Sagdeev
potential formalism and a comparison between this approach
(corresponding to strong nonlinearity), and the the reductive
perturbation approach (weak nonlinearity and weak dispesion
$k^{3}\ll 1$) leading to the KdV equation was made. The theory of
nonlinear waves in an nondegenerate ultrarelativistic plasma in
the short-wavelength limit $k\gg 1$, where the linear dispersion
has an exponential character $\omega\sim k\exp (-k^{2})$ (known in
physics as the so-called dispersion of "zero sound"
\cite{Landau_Lifshitz9}), is fully absent. Recently, a novel
nonlinear evolution equation with such dispersion and quadratic
nonlinearity was derived by using kinetic equation for the Wigner
function in Ref.~\cite{Lashkin2020} for short-wavelength
longitudinal waves in a nonrelativistic fully degenerate
electron-ion quantum plasma. It was shown \cite{Lashkin2020} that
despite the specific nature of the dispersion which has no
counterpart in classic nonrelativistic plasmas, balance between
the weak dispersion $k\gg 1$ and weak quadratic nonlinearity lead
to the formation of solitons and the collisions between three
solitons are elastic. The goal of the present Brief Communication
is to derive a similar evolution equation describing
short-wavelength nonlinear waves in a classic nondegenerate
electron-positron-ion plasma with ultrarelativistic Maxwellian
distribution function for electrons and positrons, and
nonrelativistic one for ions. Unlike the work \cite{Lashkin2020},
all three species of particles (electrons, positrons and ions)
should be present. Electrons and positrons account for the linear
dispersion, while the ions do for the nonlinearity since the
electron and positron nonlinear contributions are canceled each
other. In addition, we numerically demonstrate that collisions
between even four solitons are elastic.

Dielectric functions and dispersion relations of ultrarelativistic
plasmas was first obtained by Silin in Ref.~\cite{Silin1960}.
Expression for the longitudinal $\varepsilon_{L}$ dielectric
permittivity of an unmagnetized isotropic ultrarelativistic plasma
is \cite{Silin1960,Melrose,Landau_Lifshit10}
\begin{gather}
\varepsilon_{L}
(\omega,\mathbf{k})=1+\frac{3\omega_{p}^{2}}{k^{2}c^{2}}\left[1+
\frac{\omega}{2kc}\ln\left|\frac{\omega-kc}{\omega+kc}\right|\right]
 \nonumber \\
+i\pi\frac{\omega}{2kc}\theta (k^{2}c^{2}-\omega^{2}),
\label{permittivity}
\end{gather}
where $\theta(x)$ is the Heaviside step function, $\omega$ and
$\mathbf{k}$ are the frequency and wave vector respectively,
$k=|\mathbf{k}|$, $c$ is the speed of light, and
$\omega_{p}^{2}=\omega^{2}_{pe}+\omega^{2}_{pp}$, where
$\omega_{p\alpha}$ is the plasma frequency of the particles of
species $\alpha=e,p$ (electrons, positrons) determined by
\begin{equation}
\omega_{p\alpha}=\sqrt{4\pi e^{2}c^{2}n_{0\alpha}/3T_{\alpha}}.
\end{equation}
Here, $e$ is the elementary charge, $n_{0\alpha}$ and $T_{\alpha}$
are the equilibrium plasma density and the particle temperature of
species $\alpha$ respectively. Analytical expressions for the wave
dispersion can be obtained from the dispersion equation
$\varepsilon_{L} (\omega,\mathbf{k})=0$ in the two limiting cases
\cite{Silin1960, Melrose,Landau_Lifshit10}. In the long-wave limit
$\omega_{p}\gg kc$, the dispersion relation for longitudinal waves
is
\begin{equation}
\label{displong}
\omega_{\mathbf{k}}=\omega_{p}\left(1+\frac{3}{10}\frac{k^{2}c^{2}}{\omega_{p}^{2}}\right).
\end{equation}
In the opposite case, i. e. in the short-wave limit $\omega_{p}\ll
kc$, the dispersion relation has the form
\begin{equation}
\label{dispersion}
\omega_{\mathbf{k}}=kc\left[1+2\exp\left(-\frac{2}{3}\frac{k^{2}c^{2}}
{\omega_{p}^{2}}-2\right)\right].
\end{equation}
As is seen from Eq. (\ref{permittivity}), the Landau damping is
absent in both cases since $\omega/k>c$. Equation (\ref{displong})
is similar to the dispersion of Langmuir waves in classical
nonrelativistic electron-ion plasmas. In particular, balance
between the dispersion  Eq. (\ref{displong}) and cubic
nonlinearity results in ultrarelativistic Langmuir soliton
\cite{Mofiz1992,Mofiz1993,Liu2012}. The dispersion
(\ref{dispersion})  has no counterpart in nonrelativistic
classical plasmas. However, the dispersion relation Eq.
(\ref{dispersion}) is the same as in a nonrelativistic degenerate
quantum electron-ion plasma
\cite{Landau_Lifshit10,Melrose,Klimontovich1960} in the limits
$T_{e}\rightarrow 0$, $\hbar k/m\rightarrow 0$, and $\omega_{p}\ll
kv_{F}$ with the replacement $c\rightarrow v_{F}$, where
$v_{F}=\hbar (3\pi^{2}n_{0})^{1/3}/m$ is the electron Fermi speed
and $\omega_{p}\rightarrow \sqrt{4\pi e^{2}n_{0}/m_{e}}$ is the
electron Langmuir frequency, where $m_{e}$ is the electron mass
and $\hbar$ is the Planck constant divided by $2\pi$. For such a
nonrelativistic degenerate quantum  plasma, a nonlinear evolution
equation with this type of dispersion was obtained in
Ref.~\cite{Lashkin2020}. Here we derive a similar nonlinear
equation for an  ultrarelativistic nondegenerate
electron-positron-ion plasma. In contrast to
Ref.~\cite{Lashkin2020}, on the one hand, we completely neglect
quantum effects, so that all plasma particles are nondegenerate
and the corresponding condition has the form $T_{\alpha} \gg
\hbar^{2}n^{2/3}_{\alpha}/m_{\alpha}$. On the other hand, the
temperatures of electrons and positrons are so high that
$T_{e,p}\gg m_{e}c^{2}$, while ions are assumed to be
nonrelativistic $T_{i}\ll m_{i}c^{2}$, where $m_{i}$ is the ion
mass.

Obtaining of a nonlinear equation with the dispersion Eq.
(\ref{dispersion}) requires an essentially kinetic description (in
Ref.~\cite{Lashkin2020}, the kinetic equation for the Wigner
function was used). In the kinetic theory, the response of a
plasma to longitudinal (i.e. electrostatic) wave fields is
described by the linear response and a hierarchy of nonlinear
susceptibilities
\cite{Kadomtsev1965,Silin1965,Sitenko,Tsytovich1995,Yoon2005}. For
an isotropic unmagnetized relativistic plasma, general expressions
for the quadratic and cubic nonlinear response tensors were
obtained in Ref. \cite{Melrose}. We consider a plasma without an
external magnetic field. Laboratory ultrarelativistic plasmas
created by extremely high laser fields are unmagnetized. As for
the ultrarelativistic plasma of astrophysical objects, for
example, pulsars, we note that ultrarelativistic electrons and
positrons forming the pulsar plasma are forced to have a
one-dimensional motion along the extremely strong pulsar magnetic
field, due to fast radiative losses of perpendicular momentum.
However, electron-positron plasmas in early universe and inside
the gamma-ray burst fireball at the initial phase of its expansion
are likely to be unmagnetized \cite{Weinberg_book,Tajima1990}.
With this in mind, for generality we consider the
three-dimensional case for the moment. Further, when considering
the one-dimensional case, in the subsequent equations  $k$ means
the wave number along the external magnetic field. Throughout this
paper we use the notation
\begin{equation}
\sum_{q=q_{1}+q_{2}} \cdots \rightarrow\int \cdots \,\delta
(q-q_{1}-q_{2})\frac{dq_{1}}{(2\pi)^{4}}\frac{dq_{2}}{(2\pi)^{4}},
\end{equation}
where $q=(\mathbf{k},\omega)$. Neglecting the collision integral,
the kinetic Vlasov equation in the momentum space can be written
as
\begin{gather}
 (\omega-\mathbf{k}\cdot\mathbf{v})
f_{q}(\mathbf{p})+e_{\alpha}\varphi_{q}\mathbf{k}\cdot\frac{\partial
f^{(0)}_{\alpha}(\mathbf{p})}{\partial\mathbf{p}}  \nonumber \\
+e_{\alpha}\sum_{q=q_{1}+q_{2}}\varphi_{q_{1}}
\mathbf{k}_{1}\cdot\frac{\partial
 f_{q_{2}}(\mathbf{p})}{\partial \mathbf{p}}=0,
 \label{basic_kinetic}
\end{gather}
where $f_{q}(\mathbf{p})$ is the deviation of the distribution
function of each species, including ions $\alpha=i$, from the
equilibrium one $f^{(0)}_{\alpha}(\mathbf{p})$, $e_{\alpha}$ is
the charge of species, and $\varphi$ is the electrostatic
potential. The equilibrium distribution function for particles of
species $\alpha$ for isotropic unmagnetized relativistic plasma is
the J\"{u}ttner distribution function \cite{Melrose,Juttner}
\begin{equation}
\label{Juttner}
f^{(0)}_{\alpha}(\mathbf{p})=\frac{n_{0\alpha}}{4\pi
m_{\alpha}^{3}c^{3}}\frac{\rho_{\alpha}}{K_{2}(\rho_{\alpha})}\exp
(-\rho_{\alpha}\gamma_{\alpha}),
\end{equation}
where $\gamma_{\alpha}=\sqrt{1+p^{2}/(m_{\alpha}c)^{2}}$ is the
usual Lorentz factor, $K_{2}$ is the second kind modified Bessel
function of the second order, and
$\rho_{\alpha}=m_{\alpha}c^{2}/T_{\alpha}$. In the
ultrarelativistic limit $\rho_{\alpha}\ll 1$ for electrons and
positrons ($\alpha=e,p$) one has
\begin{equation}
\label{ultra_maxwell}
f^{(0)}_{\alpha}(\mathbf{p})=\frac{n_{0\alpha}c^{3}}{8\pi
T_{\alpha}^{3}}\exp (-cp/T_{\alpha}),
\end{equation}
while for nonrelativistic ($\rho_{i}\gg 1$) ions, the usual
Maxwellian distribution follows from  Eq. (\ref{Juttner}). For
ultrarelativistic electrons and positrons we have
$\mathbf{v}=\mathbf{p}c/p$ and for nonrelativistic ions
$\mathbf{v}=\mathbf{p}/m_{i}$. The distribution function
$f^{(0)}_{\alpha}(\mathbf{p})$ is normalized to the equilibrium
plasma density of each species, $\int
f^{(0)}_{\alpha}(\mathbf{p})d\mathbf{p}=n_{0\alpha}$.

We present the function $f_{q}(\mathbf{p})$ as a series in powers
of the field strength
\begin{equation}
\label{series}
f_{q}(\mathbf{p})=\sum_{n=1}^{\infty}f_{q}^{(n)}(\mathbf{p}).
\end{equation}
In the linear approximation, from Eqs. (\ref{basic_kinetic}) and
(\ref{series}) one can obtain
\begin{equation}
\label{recur0}
f_{q}^{(1)}=-\frac{e_{\alpha}\varphi_{q}}{(\omega-\mathbf{k\cdot\mathbf{v}})}\mathbf{k}\cdot\frac{\partial
f^{(0)}_{\alpha}}{\partial \mathbf{p}},
\end{equation}
and then we have the the recurrence relation
\begin{equation}
\label{recur}
f_{q}^{(n)}=-\frac{e_{\alpha}}{(\omega-\mathbf{k\cdot\mathbf{v}})}
\sum_{q=q_{1}+q_{2}}\varphi_{q_{1}}\mathbf{k}_{1}\cdot\frac{\partial
f_{q_{2}}^{(n-1)}}{\partial \mathbf{p}}.
\end{equation}
Retaining terms in Eq. (\ref{series}) up to second order in the
wave fields and substituting $f_{q}$ into the Poisson equation
\begin{equation}
k^{2}\varphi_{q}=4\pi \sum_{\alpha}e_{\alpha}\int
f_{q}(\mathbf{p})d^{3}\mathbf{p} ,
\end{equation}
where $\sum$ stands for summation over the different species, we
get
\begin{equation}
\label{nonlin_eq1} \varepsilon_{q}\varphi_{q}=\sum_{q=q_{1}+q_{2}}
V_{q_{1},q_{2}}\varphi _{q_{1}}\varphi_{q_{2}},
\end{equation}
where
\begin{equation}
\label{linear_responce} \varepsilon_{q}=1+\sum_{\alpha}\frac{4\pi
e_{\alpha}}
{k^{2}}\int\frac{\mathbf{k}}{(\omega-\mathbf{k}\cdot\mathbf{v})}
\cdot\frac{\partial f^{(0)}_{\alpha}}{\partial
\mathbf{p}}d^{3}\mathbf{p}
\end{equation}
is the linear dielectric response function, and neglecting ions
one has Eq. (\ref{permittivity}) where $f^{(0)}_{e}$ is determined
by Eq. (\ref{ultra_maxwell}). The interaction matrix element
$V_{q_{1},q_{2}}$ is determined by
\begin{gather}
V_{q_{1},q_{2}}=\sum_{\alpha}\frac{2\pi e_{\alpha}^{3}}{k^{2}}\int
\frac{\mathbf{k}_{1}\cdot\mathbf{k}_{2}}{[(\omega_{1}+\omega_{2})-(\mathbf{k}_{1}+\mathbf{k}_{2})\cdot
\mathbf{v}]} \nonumber \\
\times\frac{\partial}{\partial
\mathbf{p}}\frac{1}{(\omega_{2}-\mathbf{k}_{2}\cdot
\mathbf{v})}\cdot\frac{\partial f^{(0)}_{\alpha}}{\partial
\mathbf{p}}d^{3}\mathbf{p}+(\omega_{1},\mathbf{k}_{1}\rightleftarrows\omega_{2},\mathbf{k}_{2}).
\label{matrix1}
\end{gather}
Note that the expression Eq. (\ref{matrix1}) for the interaction
matrix element $V_{q_{1},q_{2}}$ is written in a symmetrized form.
Singularities in the denominators in Eqs. (\ref{linear_responce})
and (\ref{matrix1})  are avoided, as usual, using Landau's rule by
replacing  $\omega\rightarrow\omega+i0$. Linear Landau damping in
ultrarelativistic plasma is absent. In this paper we neglect the
nonlinear Landau damping (damping of the virtual beat wave) and
only the principal value of the corresponding integral is
understood, although the corresponding damping term can be easily
obtained  in the same way as the nonlinear Landau damping is
obtained in the kinetic derivation of the NLS equation for
Langmuir waves in classic plasmas \cite{Horton_Ichikawa1996}. It
is seen that the electron and positron terms in Eq.
(\ref{matrix1}) have the opposite signs so that the contributions
from electrons and positrons may cancel each other (complete
mutual cancelation occurs if $\omega_{pe}=\omega_{pp}$). For a
pure electron-positron plasma in the thermal equilibrium
($T_{e,p}=T$), quadratic nonlinearity vanishes identically. The
situation changes drastically if ions are present. We write the
interaction matrix element Eq. (\ref{matrix1}) as
$V_{q_{1},q_{2}}=V_{q_{1},q_{2}}^{(i)}+V_{q_{1},q_{2}}^{(e,p)}$,
where the first term corresponds to the ion contribution, and the
second to electrons and positrons. For nonrelativistic ions, after
two partial integrations in Eq. (\ref{matrix1}) one can write
\begin{gather}
V_{q_{1},q_{2}}^{(i)}= \frac{2\pi e^{3}}{k^{2}}\int
\left[\frac{2(\mathbf{k}\cdot\mathbf{k}_{1})
(\mathbf{k}\cdot\mathbf{k}_{1})}{(\omega-\mathbf{k}\cdot\mathbf{v})^{3}
(\omega_{2}-\mathbf{k}_{2}\cdot\mathbf{v})} \right.
 \nonumber \\
 \left. +\frac{\mathbf{k}\cdot\mathbf{k}_{1}k_{2}^{2}}{(\omega-
 \mathbf{k}\cdot\mathbf{v})^{2}(\omega_{2}-\mathbf{k}_{2}\cdot\mathbf{v})^{2}}\right]f^{(0)}_{\alpha}
d^{3}\mathbf{p}
+(\omega_{1},\mathbf{k}_{1}\rightleftarrows\omega_{2},\mathbf{k}_{2}),
\label{matrix2}
\end{gather}
where $\omega=\omega_{1}+\omega_{2}$ and
$\mathbf{k}=\mathbf{k}_{1}+\mathbf{k}_{2}$. For ultrarelativistic
electrons and positrons with $\mathbf{v}=\mathbf{p}c/p$ in an
isotropic plasma we rewrite Eq. (\ref{matrix1}) as
\begin{gather}
V_{q_{1},q_{2}}^{(e,p)}=\sum_{\alpha=e,p}\frac{2\pi
e_{\alpha}^{3}}{k^{2}}\int
\frac{k_{1}k_{2}\cos^{2}\theta}{[(\omega_{1}+\omega_{2})-(k_{1}+k_{2})c\cos\theta]} \nonumber \\
\times\frac{\partial}{\partial
p}\frac{1}{(\omega_{2}-k_{2}c\cos\theta)}\frac{\partial
f^{(0)}_{\alpha}}{\partial p}d^{3}\mathbf{p}
+(\omega_{1},\mathbf{k}_{1}\rightleftarrows\omega_{2},\mathbf{k}_{2}).
\label{matrix1_cos}
\end{gather}
where $\theta$ is an angle between $\mathbf{k}$ and $\mathbf{v}$,
and $d^{3}\mathbf{p}=2\pi p^{2}\sin\theta dp d\theta$. When
calculating the nonlinear term  in Eqs. (\ref{matrix2}) and
(\ref{matrix1_cos}), we neglect the dispersion corrections
(thermal corrections) to nonlinearity which correspond to
$k_{1,2}$ in the denominators and take into account only the
leading term. In the following, we consider the one-dimensional
case $k_{x}=k$,$k_{y}=k_{z}=0$ and then from Eq. (\ref{matrix2})
we find for ions
\begin{gather}
V_{_{q_{1}},q_{2}}^{(i)}=\frac{e}{2m_{i}}\frac{\omega_{pi}^{2}}{k^{2}}\left\{\frac{2k^{2}k_{1}k_{2}}
{\omega^{3}\omega_{2}}
+\frac{kk_{1}k_{2}^{2}}{\omega^{2}\omega_{2}^{2}}
 \right.
 \nonumber \\
\left.
+(\omega_{1},k_{1}\rightleftarrows\omega_{2},k_{2})\right\}.
\label{matrix3}
\end{gather}
For electrons and positrons from Eq. (\ref{matrix1_cos}) one has
\begin{equation}
\label{matrix4} V_{_{q_{1}},q_{2}}^{(e,p)}=-\frac{2\pi
e^{3}n_{0i}c^{2}}{k^{2}T^{2}}\left\{\frac{k_{1}k_{2}}
{\omega\omega_{2}}
+(\omega_{1},k_{1}\rightleftarrows\omega_{2},k_{2})\right\}.
\end{equation}
Comparing Eqs. (\ref{matrix3}) and (\ref{matrix4}) one can see
that the ion contribution in the nonlinearity is negligible.
Essentially, that the wave dispersion in Eq. (\ref{dispersion})
has an acoustic type and in the leading term satisfies the
three-wave resonance condition
\begin{equation}
\label{resonance} \omega_{k}=\omega_{k_{1}}+\omega_{k_{2}}, \,
k=k_{1}+k_{2}.
\end{equation}
In particular, this means that this condition, together with a
quadratic nonlinearity, ensures the validity of the successive
approximation in Eq. (\ref{series}), and this is equivalent
\cite{Sitenko,Sitenko1973} to the multi-time-scale perturbation
expansion, i. e. the secular terms are removed. Thus, from Eqs.
(\ref{dispersion}), (\ref{matrix4}) and (\ref{resonance}) we have
\begin{equation}
\label{matrix5} V_{_{q_{1}},q_{2}}=-\frac{2\pi
e^{3}n_{0i}}{k^{2}T^{2}}.
\end{equation}
Expanding $\varepsilon(\omega,\mathbf{k})$ in Eq.
(\ref{permittivity}) near the eigenmode $\omega_{\mathbf{k}}$
determined by Eq. (\ref{dispersion}) yields
\begin{equation}
\label{expansion} \varepsilon_{q}=\varepsilon
(\omega_{\mathbf{k}})+\varepsilon^{\prime}(\omega_{\mathbf{k}})(\omega-\omega_{\mathbf{k}})
\end{equation}
where $\varepsilon^{\prime}(\omega_{\mathbf{k}})=\partial
\varepsilon
(\omega)/\partial\omega\mid_{\omega=\omega_{\mathbf{k}}}$  and
from Eq. (\ref{permittivity}) one can obtain
\begin{equation}
\label{epsilon_der}
\varepsilon^{\prime}(\omega_{\mathbf{k}})=\frac{3\omega_{p}^{2}}{4k^{3}c^{3}}
\exp\left(\frac{2}{3}\frac{k^{2}c^{2}}{\omega^{2}_{p}}+2\right).
\end{equation}
After inserting Eq. (\ref{expansion}) into Eq. (\ref{nonlin_eq1})
we find
\begin{equation}
\label{main2}
(\omega-\omega_{\mathbf{k}})\varphi_{q}=\frac{1}{\varepsilon^{\prime}(\omega_{\mathbf{k}})}
\sum_{q=q_{1}+q_{2}}V_{q_{1},q_{2}}\varphi
_{q_{1}}\varphi_{q_{2}}.
\end{equation}
\begin{figure*}
\includegraphics[width=5.2in]{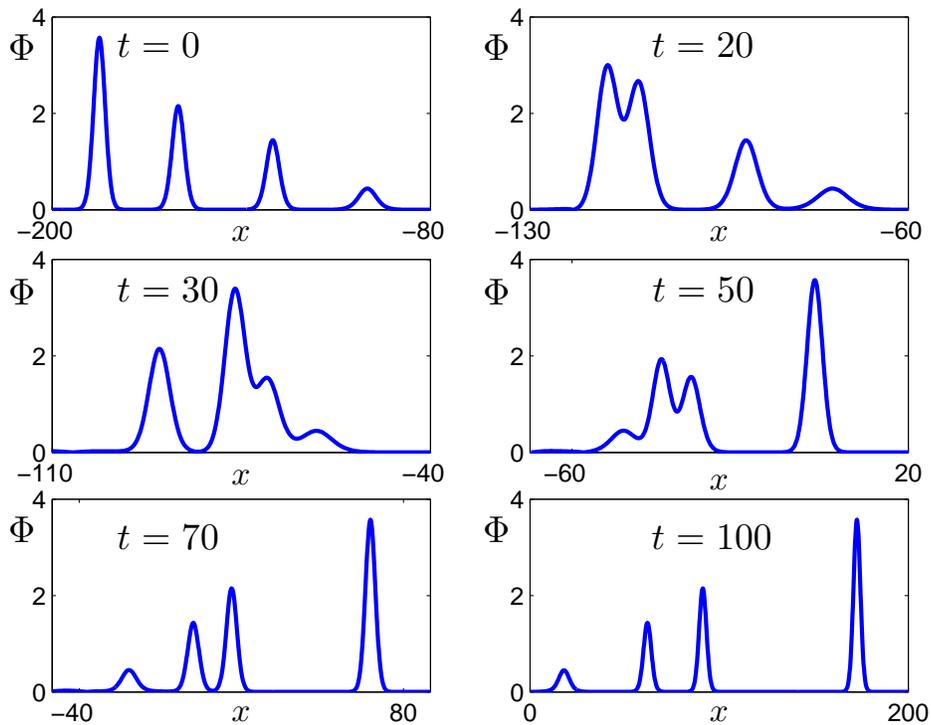}
\caption{\label{fig2} A typical example of an elastic collision
between four  solitons, with the velocities $v_{1}=3.5$,
$v_{2}=2.5$, $v_{3}=2$ and $v_{4}=1.3$. The corresponding initial
locations at the moment $t=0$ are $x_{1}=-185$, $x_{2}=-160$,
$x_{3}=-130$ and $x_{4}=-100$.}
\end{figure*}
Then, substituting Eqs. (\ref{matrix5}) and (\ref{epsilon_der})
into Eq. (\ref{main2}), and introducing the slow time scale
$\Omega=\omega-kc$ which balances the dispersion in Eq.
(\ref{dispersion}) (compare, for example, kinetic derivation of
the KdV equation in Ref.  \cite{Scorich2010}), we finally get
\begin{gather}
\left[\Omega -
2kc\exp\left(-\frac{2}{3}\frac{k^{2}c^{2}}{\omega^{2}_{p}}-2\right)\right]\varphi_{k}
\nonumber \\
=-\frac{2ecn_{0i}}{T(n_{0e}+n_{0p})}k
\exp\left(-\frac{2}{3}\frac{k^{2}c^{2}}{\omega^{2}_{p}}-2\right)
\sum_{q=q_{1}+q_{2}}\varphi_{q_{1}}\varphi_{q_{2}}. \label{main3}
\end{gather}
After rescaling
\begin{equation}
k\rightarrow \frac{c}{\omega_{p}}k, \, \Omega\rightarrow
\frac{\exp (2)}{2\omega_{p}}\Omega, \, \Phi\rightarrow
-\frac{en_{0i}}{T(n_{0e}+n_{0p})}\varphi,
\end{equation}
equation (\ref{main3}) can be written in the dimensionless form
\begin{equation}
\label{main33} \left[\Omega -
k\exp\left(-\frac{2}{3}k^{2}\right)\right]\Phi_{q} =k
\exp\left(-\frac{2}{3}k^{2}\right)
\sum_{q=q_{1}+q_{2}}\Phi_{q_{1}}\Phi_{q_{2}}.
\end{equation}
By introducing the operator $\hat{L}$ acting in the physical space
as
\begin{equation}
\hat{L}f(x)=\int ik\exp (-2k^{2}/3)\mathrm{e}^{-ikx}
\hat{f}(k)\,dk,
\end{equation}
where $f(x)$ is an arbitrary function and $\hat{f}(k)$ is its
Fourier transform, and using the convolution identity
\begin{equation}
\label{convol} (fg)_{k}=\int\hat{f}_{k_{1}}\hat{g}_{k_{2}}\delta
(k-k_{1}-k_{2})\,dk_{1}dk_{2}.
\end{equation}
one can write Eq. (\ref{main33}) in the physical space as
\begin{equation}
\label{main4}
\partial_{t}\Phi +\hat{L}\Phi +\hat{L}\Phi^{2}=0,
\end{equation}
so that the nonlinearity has a nonlocal character. Note also that
in the considered short-wavelength case $k>1$,  Eq. (\ref{main33})
can not be simplified by any expansion in $k$. Equation
(\ref{main4}) outwardly coincides with the equation obtained
earlier in Ref.~\cite{Lashkin2020} for the case of a
nonrelativistic degenerate quantum plasma using the kinetic
equation for the Wigner function, but the physical meaning of its
coefficients in dimensional form is completely different.

Stationary traveling soliton solutions of Eq. (\ref{main4}) of the
form $\Phi(x,t)=\Phi(x-vt)$, where $v$ is the velocity of
propagation in the $x$ direction was obtained numerically in Ref.
\cite{Lashkin2020}. Soliton solutions exist provided by the
condition $v>1$. In physical variables this means that the soliton
velocity should satisfy the condition $v>2\exp (-2)c\sim 0.27c$.
In reality, the soliton velocity can not be superluminal as for
any physical object, so that in the dimensionless variables the
restriction above is $v<\sim 3.7$. Note that for the group
velocity $v_{gr}=\partial\omega/\partial k$ of linear waves with
dispersion Eq. (\ref{dispersion}) we have $v_{gr}<c$. The time
evolution of the solitons under their collisions was studied in
Ref. ~\cite{Lashkin2020}, where it was shown that collisions
between two and three solitons of Eq. (\ref{main4}) are fully
elastic. In the present paper, we numerically solve the nonlinear
equation Eq. (\ref{main4}) with the initial conditions given by a
superposition of $N=4$ soliton solutions
\begin{equation}
\Phi(x,t)=\sum_{i=1}^{N}\Phi_{i}(x-x_{i},t)
\end{equation}
at the time $t=0$, where $\Phi_{i}$ correspond numerically found
(up to machine accuracy) soliton solutions with essentially
different velocities $v_{i}$. An example of the elastic collision
between four solitons with the velocities $v_{1}=3.5$,
$v_{2}=2.5$, $v_{3}=2$ and $v_{4}=1.3$ is shown in
Fig.~\ref{fig2}. In particular, it can be seen that at the times
$t=30$ and $t=50$ in the  Fig.~\ref{fig2} the three solitons
undergo strong distortion simultaneously so that two distant
solitons feel each other through an intermediate soliton -- this
is a typical many-soliton effect (at the time $t=30$ even all four
solitons feel each other). Then, the solitons fully reconstruct
their initial form without any emitting wakes of radiation
($t=100$), resulting only in phase shifts. The overall picture
closely resembles the elastic soliton collisions in the integrable
models \cite{Zakharov}. The elastic collisions between solitons
might suggest that equation (\ref{main4}) has exact $N$-soliton
solutions and is completely integrable just like for KdV equation
and others \cite{Zakharov,Ablowitz1987,Tahtadjan1987} and can be
solved by the inverse scattering transform (IST) method, but, as
was pointed out in Ref. \cite{Lashkin2020}, this is most likely
not the case. The fact is that in the IST approach there exists a
relationship between some function $\hat{\omega}(\lambda)$, where
$\lambda$ is the spectral parameter and $\omega (k)$ is the
dispersion relation of the corresponding linearized equation
\cite{Ablowitz1987}. In all known cases $\hat{\omega}(\lambda)$ is
the rational function of $\lambda$ though the associated spectral
problem may involve meromorphic functions (like the elliptic
Jacobi functions, as in the case of the Landau-Lifshitz equation
\cite{Tahtadjan1987}) of the spectral parameter $\lambda$.
However, in addition we would like to make the following remark.
In the Hirota bilinearization method, equations admitting
$N$-soliton solutions are written in the so-called bilinearization
form \cite{Bishop2}. In this case, the corresponding function of
the so-called Hirota operators $D_{t}$ and $D_{x}$  is a
polynomial (this reflects the character of linear dispersion and
takes place for all known equations considered in the Hirota
method) \textsl{or an exponential} (to our knowledge, such
equations have not been considered) \cite{Bishop2}. An interesting
question arises whether Eq. (\ref{main4}) can be written in the
bilinearization form such as (in the Hirota notations), for
example, $D_{x}[D_{t}+D_{x}\exp (-D_{x}^{2})]F\cdot F=0$ or
something like this (for the KdV equation, the bilinearization
form is $D_{x}(D_{t}+D_{x}^{3})F\cdot F=0$)?

Note also that, for illustrative purposes, we also considered Eq.
(\ref{main4}) with the replacement of the nonlocal nonlinearity by
the usual local nonlinearity $\Phi\partial_{x}\Phi$ of the KdV
equation. It turned out that soliton solutions exist in a rather
narrow range of soliton velocities and amplitudes (in particular,
there are velocity limitations both from below and from above).
Moreover, collisions of two solitons are not elastic in that
model.

In summary, we have derived the nonlinear evolution equation
governing dynamics of the short-wavelength longitudinal waves in
the ultrarelativistic plasma. In contrast to the work
\cite{Lashkin2020}, where a similar equation was derived for a
nonrelativistic fully degenerate quantum electron-ion plasma, in
our case all three species of particles (electrons, positrons and
ions) must be present. Electrons and positrons account for the
linear dispersion, while the ions do for the nonlinearity. We have
demonstrated that the collisions between even four solitons are
fully elastic resulting only in phase shifts.

\section*{DATA AVAILABILITY}
Data sharing is not applicable to this article as no new data were
created or analyzed in this study.

\bibliography{Ultra_POP_resub}

\end{document}